\documentstyle[12pt,newpasp,twoside, epsf]{article}

\newcommand{\ie}{{\em i.e.},\ }
\newcommand{\eg}{{\em e.g.},\ }
\newcommand{\etal}{{\em et al.}\ }

\newcommand{\etc}{{\em etc.}\ }

\def\edcomment#1{\iffalse\marginpar{\raggedright\sl#1\/}\else\relax\fi}
\reversemarginpar

\begin{document}
\title{Exploring the Multi-Wavelength, Low Surface Brightness Universe}
\author{R.J. Brunner, S.G. Djorgovski, R.R. Gal, A.A. Mahabal}
\affil{Department of Astronomy, California Institute of Technology, Pasadena, CA, 91125}
\author{S.C. Odewahn}
\affil{Department of Physics \& Astronomy, Arizona State University, Tempe, AZ, 85287}

\begin{abstract}

Our current understanding of the low surface brightness universe is
quite incomplete, not only in the optical, but also in other
wavelength regimes.  As a demonstration of the type of science which
is facilitated by a virtual observatory, we have undertaken a project
utilizing both images and catalogs to explore the multi-wavelength,
low surface brightness universe. Here, we present some initial results
of this project. Our techniques are complimentary to normal data
reduction pipeline techniques in that we focus on the diffuse emission
that is ignored or removed by more traditional algorithms. This
requires a spatial filtering which must account for objects of
interest, in addition to observational artifacts (\eg bright stellar
halos). With this work we are exploring the intersection of the
catalog and image domains in order to maximize the scientific
information we can extract from the federation of large survey data.

\end{abstract}

\section{Introduction}

Looking at large scale images (\ie several degrees or more), one is
immediately drawn to the high density of small galaxies, especially at
high Galactic latitude, which are often strongly
clustered. Interestingly enough, the vast majority of these galaxies
have optical surface brightness distributions that are nearly
identical to the terrestrial sky (Freeman 1970, Disney 1976). This
interesting point, unless it is the manifestation of a cosmic
coincidence, is most easily explained by accepting that current
surveys suffer from an implicit surface brightness selection effect
(see Disney 1998 for a stimulating discussion). As a result, untold
numbers of galaxies remain uncatalogued with many important
consequences.

For example, while the theoretical predictions of models of
hierarchical structure formation have been successful in predicting
the observed properties of the high redshift universe, they tend to
over-predict the number of observed local group galaxies (Klypin \etal
1999). This situation can be viewed as either a failure of the models,
or a failure of the observations, possibly due to selection
effects. In addition, galaxies which have low surface brightness (LSB)
distributions have enormous cosmological implications (\eg Impey \&
Bothun 1997), yet, they are relatively unexplored, primarily due to
the intrinsic selection effects in finding them. For example, LSB
galaxies constitute an unknown fraction of mass (both baryonic and
dark matter) which must be accounted for when determining $\Omega$ or
$\Lambda$.

In order to address these questions, we have initiated a project to
reprocess the Digitized Palomar Observatory Sky Survey (DPOSS,
Djorgovski \etal 1998) optical photographic plate data in effort to
find previously unknown, low surface brightness objects (Brunner \etal
2001). Our techniques are complimentary to normal data reduction
pipeline techniques in that we focus on the diffuse emission that is
ignored or removed by more traditional algorithms (see also,
Armandroff \etal 1998). This requires a spatial filtering which must
account for objects of interest, in addition to observational
artifacts (\eg bright stellar halos).  As part of this project, we
have developed a novel background enhancement technique to look for
new low surface brightness sources (see Figure 1 for a
demonstration). Additional aspects affecting low surface brightness
galaxy research in the context of a virtual observatory are addressed
elsewhere (see, \eg Schombert, J. in this volume).

\begin{figure}[!htb]
\plotfiddle{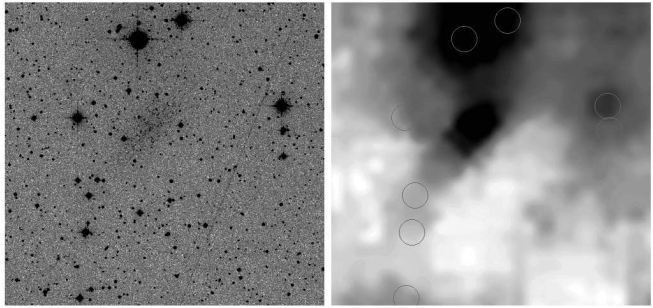}{1.5 in}{0}{100}{100}{-120}{0}
\caption{The left image is an approximately 17 arcminute square cutout 
of the DPOSS F plate image containing the dwarf Spheroidal Andromeda
III. The image on the right is the background enhanced image generated
by our software pipeline. The elongated object in the background
image, which is Andromeda III, is clearly detected via this
technique.}
\end{figure}

\section{The Technique}

Overall, the software pipeline we developed utilizes publicly
available software tools, such as SExtractor (Bertin \& Arnouts 1996),
to generate our final candidate lists. With this technique, we can
easily apply the same process to additional datasets, either
individually or jointly, in a full multi-wavelength exploration of
parameter space.

Briefly, our software pipeline performs the following steps.

\begin{itemize}

\item Pull raw DPOSS $F$ and $J$ plate scan footprints off on-line storage.

\item Mosaic full plate images.

\item Apply Vignetting correction to full plate mosaics.

\item Process the full plates to produce a background map, 
an object map, and a bright star catalog.

\item Process the background map using optimized convolution kernel, using the 
object map as a pixel mask, to detect background variations.

\item Eliminate bright stellar halos using the bright star catalog.

\item Combine candidate lists from J and F plates to remove candidates 
that arise from individual plate defects.

\item Visually classify candidates according to assigned classes (\eg 
planetary nebula, dwarf spheroidal galaxy, low surface brightness
galaxy, \etc)

\end{itemize}

In the past, non local group LSB galaxies were identified by visually
inspecting POSS-I or POSS-II sky survey plates (\eg Schombert \etal
1995). Previously, this process, particularly automated approaches,
was hampered by the unknown vignetting corrections, the plate
mosaicing process, as well as the sheer amount of data that needs to
be explored. As part of the Digital Sky project, a technology
demonstrator for a future National Virtual Observatory, these types of
problems are being tackled, and algorithmic solutions are close to
being implemented. We, therefore, plan to extend the automated low
surface brightness survey to target the unexplored population of low
surface brightness disk galaxies (see Figure 2 for a
demonstration). 

The technique we have developed can be easily applied to other large
imaging surveys, \eg the SDSS and 2MASS surveys, in an effort to
further explore the low surface brightness universe. In addition, this
work can be naturally extended to included additional wavelength
information from supplemental surveys in order to improve the source
classification (see, \eg Brunner, R. this volume) since Astronomical
objects have different source characteristics, including surface
brightness and morphology, at different wavelengths. Other,
complimentary projects are also being applied to the DPOSS data in
order to extract the maximal amount of information from this
photographic plate data (see, \eg Sabatini \etal 2000).

\begin{figure}[!htb]
\plotfiddle{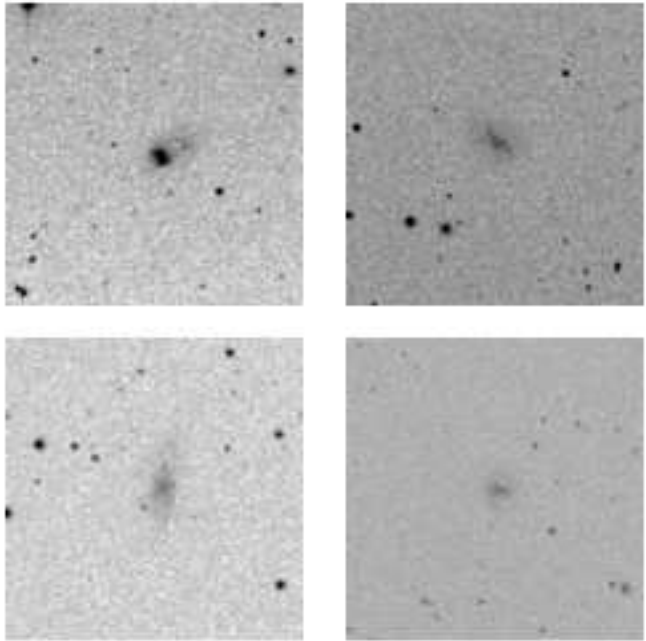}{3.5 in}{0}{100}{100}{-120}{0}
\caption{Cutouts of low surface brightness galaxies first detected visually
from POSS-II plates by J. Schombert, and later detected in HI at
Arecibo.}
\end{figure}

\section{Contaminants}

Sometimes the background enhancement procedure picks up objects which,
while interesting in their own right, are not the objects of primary
interest. Primarily these objects are either planetary nebulae,
stellar clusters, interacting galaxies, or galaxy clusters (see Figure
3 for a montage). All of these candidates are flagged due to the
presence of low surface brightness features: the nebula itself, the
combined stellar halos, the tidal interactions, and the cD envelope,
respectively.

\begin{figure}[!htb]
\plotfiddle{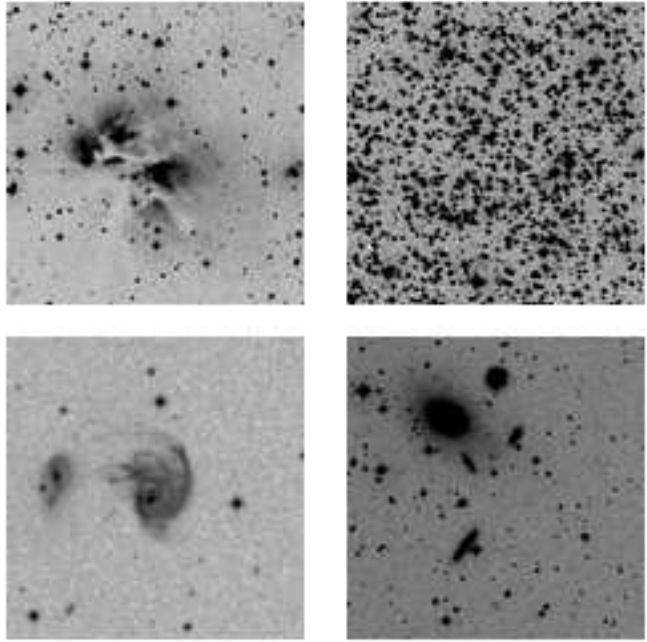}{3.5 in}{0}{100}{100}{-120}{0}
\caption{Examples of some diverse types of contaminants generated by 
the LSB software pipeline. The upper left figure is a Planetary
Nebula, the upper right figure is an open cluster. The figure on the
lower left is a pair of interacting galaxies, while the figure on the
lower right is a galaxy cluster.}
\end{figure}

\acknowledgements

This work was made possible in part through the NPACI sponsored
Digital Sky project and a generous equipment grant from SUN
Microsystems. RJB would like to acknowledge the generous support of
the Fullam Award for facilitating this project. Access to the DPOSS
image data stored on the HPSS, located at the California Institute of
Technology, was provided by the Center for Advanced Computing
Research. The processing of the DPOSS data was supported by a generous
gift from the Norris foundation, and by other private donors.

\end{document}